\documentclass[12pt]{article}

\usepackage[normalem]{ulem}

\usepackage{xcolor}
\usepackage[english]{babel}
\usepackage[T1]{fontenc}
\usepackage[utf8]{inputenc}
\usepackage{authblk}
\usepackage{mathtools}
\usepackage{slashed}
\usepackage{amsmath,amssymb}
\usepackage{amsfonts}
\usepackage{enumitem}
\usepackage{graphicx,color,xcolor}
\usepackage{cite}
\usepackage{tabularx}
\usepackage{subcaption}
\usepackage{hyperref}
\hypersetup{
    colorlinks=true,
    linkcolor=blue,
    filecolor=magenta,      
    urlcolor=cyan,
}
\usepackage[capitalize]{cleveref}

\numberwithin{equation}{section} 

\definecolor{refcol}{rgb}{0.9,0.1,0.1}
\hypersetup{colorlinks=true,linkcolor=blue,citecolor=refcol,urlcolor=cyan,linktocpage}

\graphicspath{{Plots/}}

\begin{document}
%%%%%%%%%%%%%%%%%%%%%%%%%%%%%%%%%%%%%%%%%%
\begin{titlepage}
\thispagestyle{empty}

%\title{{\Huge\bf Exact thermal correlator of $CFT$ with chemical potential}}

\title{
{\Huge\bf Exact thermal correlators of holographic $CFT$s }
}

%\bigskip\bigskip\bigskip\bigskip\bigskip
\vfill

\author{
	{\bf Atanu Bhatta$^a$}\thanks{{\tt atanu.bhatta@wits.ac.za}},  
	{\bf Taniya Mandal$^a$}\thanks{{\tt taniya.mandal@wits.ac.za}}
	\smallskip\hfill\\      	
	\small{
		$^a${\it National Institute of Theoretical and Computational Sciences}\\
		{\it School of Physics and Mandelstam Institute for Theoretical Physics}\\
		{\it University of the Witwatersrand, Johannesburg, Wits 2050, South Africa.}
			}
		}

%\bigskip\bigskip\bigskip\bigskip
\vfill

\date{
\vspace{1cm}
\begin{quote}
\centerline{{\bf Abstract}}
{\small
We compute the exact retarded Green's functions in thermal $CFT$s with chemical potential and angular momenta using holography respectively. We consider the field equations satisfied by the quasi-normal modes in both charged and rotating black holes in $AdS$ spacetime and mapped them to the Heun equations by appropriate changes of variables. The AGT correspondence allows us to find the connection formulae among the solutions of the Heun equations near different singularities by using the crossing relations of the five-point correlators in the Liouville $CFT$. The connection formulae associated with the boundary conditions imposed on the bulk field equations yield the exact thermal correlators in the boundary $CFT$. 
}
\end{quote}
}

%\vfill
%\leftline{{\bf Report No: }} }

%\begin{abstract}
%\end{abstract}

\end{titlepage}
%%%%%%%%%%%%%%%%%%%%%%%%%%%%%%%%%%%%%%%%%%
\thispagestyle{empty}\maketitle\vfill \eject

\tableofcontents
%\newpage

%%%%%%%%%%%%%%%%%%%%%%%%%%%%%%%%%%%%%%%%%%
\section{Introduction}
The retarded Green's functions of a finite temperature conformal field theory ($CFT$) are useful to study quantum transports \cite{Policastro:2001yc,Hartnoll:2016apf}, chaotic behaviour \cite{Grozdanov:2017ajz,Blake:2018leo} etc. of the given system. The $AdS/CFT$ correspondence~\cite{Maldacena:1997re,Gubser:1998bc,Witten:1998qj} dictates that the dual gravity theory of a thermal $CFT$ residing at the boundary of an asymptotically $AdS$ space can be described by an $AdS$ black hole~\cite{Witten:1998zw}. In this setup, the retarded Green's functions of the boundary $CFT$ can be obtained by considering quasi-normal modes in the dual bulk with appropriate boundary conditions. Specifically, one imposes in-going boundary  conditions on the quasi-normal modes near the horizon and identifies the response and source terms of the solution at the asymptotic boundary. Then the ratio of the response function to the source yields the retarded Green's function~\cite{Son:2002sd,Policastro:2002se,Nunez:2003eq}.

The wave equation satisfied by the quasi-normal modes in the bulk can be recast into the Heun equation by making appropriate changes of variables and field redefinitions~\cite{Musiri:2003rs,BarraganAmado:2018zpa,Amado:2021erf,Hortacsu:2011rr}. The Heun equation is the linear second order differential equation with four regular singular points~\cite{fiziev2015heun,ronveaux1995heun}. One can solve the Heun equation around the singular points to get different solutions. These are basically the expansions of the Heun functions around the points under considerations. 

Quasi-normal modes of black holes can also be studied via quantum Seiberg-Witten curve for $\mathcal{N}=2$, SYM theory \cite{Aminov:2020yma,Bianchi:2021xpr,Bonelli:2021uvf,Bianchi:2021mft}. One also finds the Heun equations while studying the scattering of $Kerr-AdS$ black holes using monodromy\cite{Amado:2020zsr,BarraganAmado:2021uyw,Novaes:2014lha}. The Heun equation appears in the study of surface operators in $\mathcal{N}=2, SU(2)$ supersymmetric gauge theory with flavors $N_f \le 4$ in four dimensions \cite{Nekrasov:2009rc}. Moreover, the Alday-Gaiotto-Tachikawa (AGT) correspondence relates the BPS sector of the gauge theory in the so-called $\Omega$-background to the two-dimensional Liouville $CFT$ \cite{Alday:2009aq,Alday:2009fs}. The four-point conformal block of the primary fields in the Liouville $CFT$ corresponds to the Nekrasov partition function of the gauge theory with four flavors and inserting degenerate fields into the correlator amounts in inserting the surface operators in the gauge theory \cite{Maruyoshi:2010iu}. In particular, in the semi-classical limit the BPZ equation~\cite{Belavin:1984vu} satisfied by the five-point correlator with one degenerate insertion in the Liouville theory becomes the Heun equation, while the points of primary operator insertions are identified with the singular points of the Heun equation \cite{Piatek:2017fyn}. 

One can solve for the conformal blocks around the different points of primary operator insertions with appropriate boundary conditions~\cite{Dolan:2000ut,Dolan:2003hv,Dolan:2011dv}. These solutions are called the conformal blocks of different channels, namely $s, t$ and $u$ channels of the corresponding correlation function. The different channels are related to each other via crossing relations. Thus using crossing relations, the solution of the Heun equation near a given singular point admits an expansion in terms of the solutions around the other singularities. These are called connection formulas between the Heun functions around different singular points~\cite{Bonelli:2022ten}. 

Finally, from the connection formula of the Heun function associated with the in-going solutions of the bulk quasi-normal mode and that associated with the asymptotic behaviour of the quasi-normal mode one can read the source and the response function and hence obtain the retarded Green's function for the theory under consideration. In this way, a closed form expression of the exact Green's function has been obtained recently for the $AdS$-Schwarzschild background which describes a finite temperature $CFT$ at the boundary~\cite{Dodelson:2022yvn}.  Earlier, attempts to find thermal two-point correlators of holographic $CFT$s have been made using Geodesic approximation method, in the black brane background in the large conformal dimension limit  \cite{Rodriguez-Gomez:2021pfh,Rodriguez-Gomez:2021mkk,Georgiou:2022ekc}. Whereas the computation of the same using quasi-normal modes in the dual $AdS$ space and connecting with the Heun equation is exact \cite{Dodelson:2022yvn}.  

We want to generalize this idea for the charged black hole, namely the Reissner-Nordstr\"{o}m $AdS_5$ black hole, for which the boundary theory becomes a finite temperature $CFT_4$ with a chemical potential. The chemical potential is related to the charge of the black hole sitting in the bulk. We consider the charged scalar quasi-normal modes in the bulk, and obtain an expression for the exact Green's function following the prescription described above. We further extend the computation of exact thermal correlator for a $CFT_4$ that is dual to a rotational $AdS$ black hole.

Organisation of this paper is as follows: In section \ref{sec:rnbh}, we discuss the computation of the exact two-point thermal correlator of a $CFT$ with chemical potential. Retarded Greens function calculation for a rotational $CFT$ has been analyzed in section \ref{sec:rotbh}. Finally, we conclude in section \ref{sec:con}.

%
%\section{Charged $AdS_5$ black hole}\label{sec:rnbh}
%\section{Exact holographic thermal correlator of a $CFT$ dual to a charged $AdS_5$ black hole}\label{sec:rnbh}
\section{$CFT$ dual to a charged $AdS_5$ black hole}\label{sec:rnbh}
In this section we compute the exact thermal correlator of a $CFT$ that is dual to a five dimensional charged black hole in asymptotically Anti-de sitter space.
The five dimensional Einstein-Maxwell theory with a negative cosmological constant is described by the following action
\begin{eqnarray}
S =\frac{1}{16\pi G_5} \int d^5x \sqrt{-g} \left(R + 12- \frac{1}{g_s^2} F_{\mu\nu}F^{\mu\nu} \right)
\end{eqnarray}
%where we have set the $AdS$ radius $L=1$. 
Solution to the equations of motion derived from the above action is called the Reissner-Nordstr\"{o}m (RN) black hole. The five dimensional  Reissner-Nordstr\"{o}m black hole in $AdS$ space is given by  
\begin{eqnarray}
    ds^2 =-f(r)dt^2+\frac{dr^2}{f(r)}+r^2 d\Omega_3^2; \qquad A = \mu \left( 1- \frac{r_+^2}{r^2}\right) \,dt.
\end{eqnarray}
Whereas the blackening factor $f(r)$ and the chemical potential $\mu$ are  
\begin{eqnarray}
    f(r) = 1-\frac{M}{r^2}+\frac{Q^2}{r^4}+\frac{r^2}{L^2}, \quad \text{and} \quad \mu = \sqrt{\frac{3}{4}} \frac{g_s Q}{r_+^2},
\end{eqnarray}
with $r_+$ being the radius of the outer horizon. The black hole parameters $M$ and $Q$ are related to the ADM mass and charge of the black hole as
\begin{eqnarray}
 M_{ADM} = \frac{3 M \Omega_3}{16 \pi G_5}; \quad \text{and} \quad Q_e = \frac{\sqrt{3}Q \Omega_3}{4\pi G_5 g_s},
\end{eqnarray}
respectively. The radius of the outer horizon $r_+$ is related to the black hole parameters as
\begin{eqnarray}
 M = r_+^2 + \frac{Q^2}{r_+^2} + \frac{r_+^4}{L^2}.
\end{eqnarray}
In terms of the outer horizon $r_+$, the blackening factor $f(r)$ can be written as  \cite{Amado:2021erf}
\begin{eqnarray}
    f(r) =\frac{(r^2-r^2_+)(r^2-r^2_-)(r^2-r^2_0)}{L^2r^4},
\end{eqnarray}
where
\begin{eqnarray}\label{rmr0}
    r_-^2 &=& \frac{L^2}{2}\left(-1-\frac{r^2_+}{L^2} + \sqrt{\left(1+\frac{r^2_+}{L^2}\right)^2+\frac{4Q^2}{L^2 r^2_+}}\right),\nonumber\\
 r_0^2 &=& \frac{L^2}{2}\left(-1-\frac{r^2_+}{L^2} - \sqrt{\left(1+\frac{r^2_+}{L^2}\right)^2+\frac{4Q^2}{L^2 r^2_+}}\right).
 \end{eqnarray}
 Temperature at the outer horizon $r_+$ is given by 
 \begin{eqnarray}\label{temp}
  T = \frac{1}{\pi}\left(\frac{r_+}{L^2}+\frac{1}{2r_+}-\frac{Q^2}{2r_+^5}\right).
 \end{eqnarray}
 Here onward, for simplicity, we set the $AdS$ radius $L=1$. In this set up, we consider a charged scalar field excitation with mass $m$ and charge $e$ on the RN-$AdS$ black hole background. The mass of the bulk scalar field is related to the conformal dimension $\Delta$ of a scalar primary operator in the boundary $CFT$ as
 \begin{eqnarray}
 m^2 = \Delta(\Delta-4).
 \end{eqnarray}
 In the charged black hole background, the Klein-Gordon equation satisfied by the charged scalar field $\phi$ reads as
\begin{eqnarray}\label{cskg}
    \frac{1}{\sqrt{-g}}D_\mu\left(\sqrt{-g} g^{\mu\nu}D_\nu\right)\phi -m^2 \phi = 0,
\end{eqnarray}
where the covariant derivative $D_\mu =\nabla_\mu - i e A_\mu$. Decomposing the scalar field in the Fourier modes as follows
\begin{eqnarray}
    \phi(t,r,\Omega) = \int d\omega \sum_{\ell, m_1, m_2} e^{-i\omega t}Y_{\ell}^{m_1,m_2}(\Omega) \psi_{\omega \ell}(r),
\end{eqnarray}
we can write the wave equation \eqref{cskg} in the following form
\begin{eqnarray}
 \Bigg(   \frac{1}{r}\frac{d}{dr}\left(r^3 f(r)\frac{d}{dr}\right)+\frac{r^2}{f(r)}
    \left(\omega-\frac{\sqrt{3}eQ}{2r^2}\right)^2-m^2 r^2-\ell(\ell+2)\Bigg)\psi_{\omega \ell}(r)=0     .
\end{eqnarray}

We want to solve the wave equation with the following boundary conditions. At the outer horizon, $r = r_+$ we impose an in-going boundary condition as
\begin{eqnarray}
\label{eq:bc@horizon}
 \psi^{in}_{\omega \ell} (r) = (r-r_+)^{-i\frac{(\omega-\omega_c) r_+^3}{2(r_+^2-r_0^2)(r_+^2-r_-^2)}} + \cdots 
\end{eqnarray}
where,
\begin{eqnarray}
 \omega_c = \frac{\sqrt{3}e Q}{2r_+^2}.
\end{eqnarray}
The solution behaves at the boundary $(r \to \infty)$ as
\begin{eqnarray}
\label{eq:bc@bdy}
 \psi^{in}_{\omega \ell} (r) = \mathcal{A}(\omega, \ell) \left(r^{\Delta -4} + \cdots \right) +\mathcal{B}(\omega, \ell) \left(r^{-\Delta} + \cdots \right)  .
\end{eqnarray}
Then the retarded Green's function can be computed as
\begin{eqnarray}
 G_R(\omega, \ell) = \frac{\mathcal{B}(\omega, \ell)}{\mathcal{A}(\omega, \ell)}.
\end{eqnarray}
\subsection{Heun equation}
Now to compute the ratio of the respose $\mathcal{B}(\omega, \ell)$ to the source $\mathcal{A}(\omega, \ell)$, we follow the technique of~\cite{Dodelson:2022yvn}. 
We adopt the following variable change
\begin{eqnarray}
 z &=& \frac{r^2-r^2_-}{r^2-r_0^2}\\
 \psi_{\omega \ell}(r) &=& z^{-\frac{1}{2}} (z-z_0)^{-\frac{1}{2}}(1-z)^{1/2} \chi_{\omega \ell}(z)
\end{eqnarray}
and recast the wave equation to the Heun equation in normal form as
\begin{eqnarray}
\label{eq:heun}
\Bigg( \partial_z^2+\frac{\frac{1}{4}-a_1^2}{(z-1)^2}-\frac{\frac{1}{2}-a_0^2-a_1^2 a_z^2+a_{\infty}^2+u}{z(z-1)}+\frac{\frac{1}{4}-a_z^2}{(z-z_0)^2}+\frac{u}{z(z-z_0)}+\frac{\frac{1}{4}-a_0^2}{z^2}\Bigg) \chi_{\omega l}(z)=0.\nonumber\\
\end{eqnarray}
The Heun equation has four singularities at $z = 0, 1, z_0$ and $\infty$, where 
\begin{eqnarray}
\label{eq:z0}
z_0 = \frac{r^2_+-r^2_-}{r^2_+-r^2_0}.
\end{eqnarray}
The outer horizon $(r=r_+)$ and the boundary $(r \to \infty)$ are mapped to $z =z_0$ and $z=1$ respectively.

We identify the parameters of the Heun equation for the charged $AdS$ black hole to be
\begin{eqnarray}\label{heunchargepara}
a_0 &=& -\frac{i r_-(r_-^2 \omega-r_+^2 \omega_c)}{2(r_-^2-r_+^2)(1+2r_-^2+r_+^2)},\\
a_z &=& \frac{ir_+^3(\omega-\omega_c)}{2(r_+^2-r_-^2)(1+r_-^2+2r_+^2)},\\
a_1 &=& \frac{\Delta-2}{2},\\
a_\infty &=& \frac{\sqrt{(1+r_-^2+r_+^2)}\left((1+r_-^2+r_+^2)\omega+r_+^2\omega_c\right)}{2(1+2r_-^2+r_+^2)(1+r_-^2+2r_+^2)},
\end{eqnarray}
and 
\begin{eqnarray}
\label{eq:u}
u &=& -\frac{l(l+2)+2(1+r_-^2+2r_+^2)+r_+^2\Delta(\Delta-4)}{4(1+2r_-^2+r_+^2)}\nonumber\\
&&+\frac{r_+^4(\omega-\omega_c)\left(r_-^2(\omega_c-3\omega)+r_+^2(\omega+\omega_c)\right)}{4(r_-^2-r_+^2)^2(1+2r_-^2+r_+^2)(1+r_-^2+2r_+^2)}.
\end{eqnarray}
 In the uncharged limit $Q\rightarrow 0\implies r_-\rightarrow 0, \omega_c\rightarrow 0$, the parameters \eqref{heunchargepara}-\eqref{eq:u} boil down to \cite{Dodelson:2022yvn}
\begin{eqnarray}\label{unchargedheunchargepara1}
	\tilde{a}_0 &=& 0,\\\label{unchargedheunchargepara2}
	\tilde{a}_z &=& \frac{ir_+\omega}{2(1+2r_+^2)},\\\label{unchargedheunchargepara3}
	\tilde{a}_1 &=& \frac{\Delta-2}{2},\\\label{unchargedheunchargepara4}
	\tilde{a}_\infty &=& \frac{\sqrt{(1+r_+^2)}\omega}{2(1+2r_+^2)},\\\label{unchargedheunchargepara5}
	\tilde{u} &=& -\frac{l(l+2)+2(1+2r_+^2)+r_+^2\Delta(\Delta-4)}{4(1+r_+^2)}\nonumber\\
	&&+\frac{r_+^2\omega ^2}{4(1+r_+^2)(1+2r_+^2)}.
\end{eqnarray}

The boundary conditions~\eqref{eq:bc@horizon},~\eqref{eq:bc@bdy} for the in-going solution take the following form at the horizon as
\begin{eqnarray}
\label{eq:zbc@horizon}
\chi_{\omega \ell}^{(z_0)} (z) = (z_0 - z)^{\frac{1}{2}-a_z} + \cdots 
\end{eqnarray}
and at the boundary as
\begin{eqnarray}
\label{eq:zbc@bdy}
\chi_{\omega \ell}^{(1)} (z)\propto \mathcal{A}(\omega, \ell) \chi^{(1), -}_{\omega \ell} (z) +\mathcal{B}(\omega, \ell) \chi^{(1), +}_{\omega \ell} (z)
\end{eqnarray}
where
\begin{eqnarray}
 \chi^{(1), \pm}_{\omega \ell} (z) = (1-z)^{1/2 \pm a_1} + \cdots 
\end{eqnarray}
\subsection{Relation to Liouville $CFT$}
The Heun equation appears in the study of conformal blocks in the Liouville $CFT$. The five-point correlation function with one degenerate operator insertion in Liouville $CFT$ satisfies the following BPZ equation 
\begin{eqnarray}
&&\left(b^{-2} \partial_z^2 + \frac{\Delta_1}{(z-1)^2} - \frac{\Delta_1 + z_0 \partial_{z_0} + \Delta_{z_0} + z\partial_z + \Delta_{2,1}+ \Delta_0 -\Delta_\infty }{z(z-1)} +\frac{\Delta_{z_0}}{(z-z_0)^2} \right. \cr 
 && \qquad \qquad \qquad + \left. \frac{z_0}{z(z-z_0)}\partial_{z_0}-\frac{1}{z}\partial_z + \frac{\Delta_0}{z^2}\right) \langle \Delta_{\infty} |V_1(1) V_{z_0}(z_0) \Phi(z) |\Delta_0 \rangle =0,
 \label{eq:bpz}
\end{eqnarray}
where $\Delta_1$ and $\Delta_{z_0}$ are the scaling dimensions of the operators $V_1$ and $V_{z_0}$ respectively and the scaling dimension of the degenerate operator $\Phi$ is given by $\Delta_{2,1} = -\frac{1}{2} - \frac{3b^2}{4}$. Since the conformal correlators admits an expansion in terms of the conformal blocks, the five-point conformal blocks also satisfy the BPZ equation. The ODE~\eqref{eq:bpz} has four regular singularities at $z=0, 1, z_0$ and $\infty$. One can solve the ODE around different singular points and obtain different five-point blocks using appropriate boundary conditions. These answers for the five-point blocks are related to each other by crossing relations.

Now we take the semi-classical limit,
\begin{eqnarray}
b \to 0, \alpha_i \to 0, b \alpha_i = a_i (\text{finite}),
\end{eqnarray}
where the momenta $\alpha_i$'s of the primary operators are related to the scaling dimensions as
\begin{eqnarray}
\Delta_i = \alpha_i (Q- \alpha_i); \qquad Q = \left(b + \frac{1}{b} \right).
\end{eqnarray}
In the semi-classical limit, the BPZ equation~\eqref{eq:bpz} reduces to the Heun equation in normal form~\eqref{eq:heun}. The operator insertion points $z =0, 1, z_0$ and $\infty$ becomes the four singular points of corresponding Heun equation and the crossing relations among the conformal blocks descend to the connection formula between the solutions of the Heun equation.

The connection formula between the solution~\eqref{eq:zbc@horizon} near the horizon $(z = z_0)$ and the solution~\eqref{eq:zbc@bdy} near the boundary $z =1$ reads~\cite{Bonelli:2022ten,Dodelson:2022yvn}
\begin{eqnarray}
\label{eq:connection}
\chi^{(z_0)}_{\omega \ell} (z) &=& \sum_{\theta'=\pm} \left(\sum_{\theta=\pm} \mathcal{M}_{-\theta}(a_{z}, a; a_0)\mathcal{M}_{(-\theta)\theta'}(a, a_1; a_{\infty}) \right. \cr  
&& \qquad \qquad \left. z_0^{\theta a}e^{-\frac{\theta}{2}\partial_a F} \right) z_0^{\frac{1}{2}-a_0-a_{z}}(1-z_0)^{a_z-a_1} e^{-\frac{1}{2}(\partial_{a_z}+\theta' \partial_{a_1})F} \chi_{\omega \ell}^{(1), \theta'}(z)\nonumber, \\
\end{eqnarray}
with
\begin{eqnarray}
\mathcal{M}_{\theta\theta'} (a_0, a_1; a_2) &=& \frac{\Gamma(-2\theta' a_1)}{\Gamma(\frac{1}{2} + \theta a_0- \theta' a_1 + a_2)} \frac{\Gamma(1+2\theta a_0)}{\Gamma(\frac{1}{2} + \theta a_0- \theta' a_1 - a_2)}.
\end{eqnarray}
Where $\theta=\pm$ are the two fusion channels allowed by the degenerate fusion rules, $a$ is defined by the Matone relation \cite{Matone:1995rx}
\begin{eqnarray}
\label{eq:matone}
u = -a^2 + a_z^2 -\frac{1}{4} + a_0^2 + z_0 \partial_{z_0} F,
\end{eqnarray}
and $F(z_0)$ is the classical conformal block, related to the conformal block without degenerate insertion as
\begin{eqnarray}
\mathcal{F}(\Delta_{0}, \Delta_1, \Delta_{z_0}, \Delta_{\infty}; \Delta; z_0) = z_0^{\Delta-\Delta_{z_0}-\Delta_0} e^{b^{-2}(F(z_0) + \mathcal{O}(b^2)} .
\end{eqnarray}
\subsection{Retarded Green's function}
We can read the ratio of the response to the source from the connection formula~\eqref{eq:connection} and write the exact retarded Green's function as 
\begin{eqnarray}
 G_R(\omega, \ell) = (1 + r_+^2)^{2a_1} e^{-\partial_{a_1}F} \frac{\sum_{\theta=\pm} \mathcal{M}_{-\theta}(a_{z}, a; a_0)\mathcal{M}_{(-\theta)+}(a, a_1; a_{\infty})  z_0^{\theta a}e^{-\frac{\theta}{2}\partial_a F} }{\sum_{\theta=\pm} \mathcal{M}_{-\theta}(a_{z}, a; a_0)\mathcal{M}_{(-\theta)-}(a, a_1; a_{\infty})  z_0^{\theta a}e^{-\frac{\theta}{2}\partial_a F}}
\end{eqnarray}
where the parameters $z_0, a_0, a_1, a_z, a_\infty$ and $u$ are written in terms of $\omega, \ell$ and the RN-$AdS$ black hole parameters $r_0$ and $r_-$ as in~\eqref{eq:z0}-\eqref{eq:u}. One can compute $a$ using the Matone relation~\eqref{eq:matone}.

\subsection{Large black hole limit}
In this section, we are considering a case of large RN-$AdS$ black hole i.e. the outer horizon radius is very much greater than the $AdS$ length, $r_+\gg L$. From \eqref{rmr0}, we see that in this limit $r_-\rightarrow 0$ whereas $r_0^2\rightarrow -(1+r_+^2)$. Then for the large black holes, the parameters \eqref{heunchargepara}-\eqref{eq:u} of the Heun equation become 
\begin{eqnarray}\label{heunlarge}
 a_0 &=& 0\\
 a_z &=& \frac{i r_+(\omega-\omega_c)}{2(1+2r_+^2)}\\
 a_1 &=& \frac{\Delta-2}{2}\\
 a_{\infty} &=& \frac{\omega(1+r_+^2)+r_+^2\omega_c}{2\sqrt{(1+r_+^2)}(1+2r_+^2)}\\\label{eq:largeu}
 u &=& -\frac{l(l+2)+2(1+2r_+^2)+r_+^2\Delta(\Delta-4)}{4(1+r_+^2)}+\frac{r_+^2(\omega^2-\omega_c^2)}{4(1+r_+^2)(1+2r_+^2)}
\end{eqnarray}

In this case, we can express the above parameters \eqref{heunlarge}-\eqref{eq:largeu} in terms of uncharged parameters \eqref{unchargedheunchargepara1}- \eqref{unchargedheunchargepara5} as follows

\begin{eqnarray}\label{correctedheunlarge}
	a_0 &=& \tilde{a}_0\\
	a_z &=& \tilde{a}_z-\frac{i r_+\omega_c}{2(1+2r_+^2)}\\
	a_1 &=& \tilde{a}_1\\
	a_{\infty} &=&\tilde{a}_\infty+\frac{r_+^2\omega_c}{2\sqrt{(1+r_+^2)}(1+2r_+^2)}\\
	u &=& \tilde{u}-\frac{r_+^2\omega_c^2}{4(1+r_+^2)(1+2r_+^2)}
\end{eqnarray}
where $\omega_c=\frac{\sqrt{3}e Q}{2r_+^2}$.

%As, in our case, $L=1$, 
For a large black hole with small chemical potential $\mu\ll1$, $\omega_c$ becomes negligible and the temperature \eqref{temp} becomes
\begin{eqnarray}\label{largetemp}
  T = \frac{r_+}{\pi}.
 \end{eqnarray}
As $r_+\gg 1$, we find a high temperature black hole in this limit. In the limit $r_+\gg 1, \mu\ll 1$, the parameters of the Heun equation are

\begin{eqnarray}\label{heunlargesmallmu}
 a_0 &=& 0\nonumber\\
 a_z &=& \frac{i r_+\omega}{2(1+2r_+^2)}\nonumber\\
 a_1 &=& \frac{\Delta-2}{2}\nonumber\\
 a_{\infty} &=& \frac{\omega\sqrt{(1+r_+^2)}}{2(1+2r_+^2)}\nonumber\\
u &=& -\frac{l(l+2)+2(1+2r_+^2)+r_+^2\Delta(\Delta-4)}{4(1+r_+^2)}+\frac{r_+^2\omega^2}{4(1+r_+^2)(1+2r_+^2)}
\end{eqnarray}
Interestingly, from \eqref{heunlargesmallmu}, one notices that the parameters of the Heun equations of the charged, large black holes in the small chemical potential, high temperature limit are same as of uncharged black holes \cite{Dodelson:2022yvn}.

%\section{Rotational $AdS$ black hole} \label{sec:rotbh}
%\section{Exact holographic thermal correlator of a $CFT$ dual to a rotational $AdS_5$ black hole}\label{sec:rotbh}
\section{ $CFT$ dual to a rotational $AdS_5$ black hole}\label{sec:rotbh}
In this section we compute the exact, thermal two-point correlator of a $CFT$ dual to a rotating black hole in $AdS_5$ in the slow rotation approximation. In particular, we consider Myers-Perry black hole solution~\cite{MYERS1986304} in $AdS$ with a single rotation parameter $\alpha$ \cite{Hawking:1998kw}. We take into account the slow rotation i.e. $\alpha \ll 1$ and the solution is linear in $\alpha$~\cite{Tattersall:2018axd}. In this limit the metric is
\begin{eqnarray}
 ds^2& =& -\left(1-\frac{2M}{r^2}+r^2\right)dt^2+\frac{dr^2}{1-\frac{2M}{r^2}+r^2}+r^2 d\theta^2+r^2 \sin^2{\theta} d\phi^2+r^2 \cos^2{\theta} d\psi^2\nonumber\\&&-2\alpha\left(\frac{2M}{r^2}-r^2\right)\sin^2{\theta}dt d\phi.
\end{eqnarray}
Temperature  and the angular velocity of the slow rotating, one-parameter rotational black hole are given by
\begin{eqnarray}
 T=\frac{1+2r_+^2}{2\pi r_+},\quad\Omega =\frac{\alpha}{r_+^2},
\end{eqnarray}
where $r_+$ is the horizon and $AdS$ length has been set to $1$. 

We aim to solve the Klein-Gordon equation of a scalar field $\Phi$ of mass $\mu$ in this background. The equation is
\begin{eqnarray}
 \frac{1}{\sqrt{-g}}\partial_\mu (\sqrt{-g} g^{\mu\nu}\partial_\nu)\Phi-\mu^2\Phi=0.
\end{eqnarray}
Expanding the field in Fourier basis, in the linear order of $\alpha$, we can write the scalar field as \cite{BarraganAmado:2018zpa}
\begin{equation}
    \Phi = \int d\omega \sum_{l,m_1,m_2} e^{-i \omega t}e^{i m_1 \phi}e^{i m_2 \psi}\mathcal{P}(\theta)\Psi_{\omega }(r).
\end{equation}
Then the scalar wave equation takes the following form
\begin{eqnarray}
&&\frac{\mathcal{P}(\theta)}{r^3}\partial_r\left(r^3 f(r)\partial_r\right)\Psi_{\omega l}(r)+\Bigg(\frac{\omega^2}{f(r)}+\frac{2\alpha m_1 \omega(1-\frac{2M}{r^4})}{f(r)}\Bigg)\mathcal{P}(\theta)\Psi_{\omega l}(r)-\frac{m_1^2}{r^2\sin^2\theta}\mathcal{P}(\theta)\Psi_{\omega l}(r)\nonumber\\
&&+\Bigg(\frac{1}{\sin( 2\theta)}\partial_\theta(\sin (2\theta)\partial_\theta)\mathcal{P}(\theta)-\frac{m_2^2}{\cos^2\theta}\mathcal{P}(\theta)\Bigg)\frac{\Psi_{\omega l}(r)}{r^2}-\mu^2 \Psi_{\omega l}(r)\mathcal{P}(\theta)=0
\end{eqnarray}
where $f(r)=1-\frac{2M}{r^2}+r^2$. Using separation of variable method, we find a radial and an angular equation as follows
\begin{eqnarray}\label{radial}
\frac{1}{r^3}\partial_r\left(r^3 f(r)\partial_r\right)\Psi_{\omega l}(r)+\Bigg(\frac{\omega^2}{f(r)}+\frac{2\alpha m_1 \omega(1-\frac{2M}{r^4})}{f(r)}\Bigg)\Psi_{\omega }(r) -\lambda \frac{\Psi_{\omega }(r)}{r^2}-\mu^2 \Psi_{\omega }(r)=0\nonumber\\
\end{eqnarray}
\begin{eqnarray}\label{angular}
m_1^2\mathcal{P}(\theta)-\sin^2\theta\Bigg(\frac{1}{\sin( 2\theta)}\partial_\theta(\sin (2\theta)\partial_\theta)\mathcal{P}(\theta)-\frac{m_2^2}{\cos^2\theta}\mathcal{P}(\theta)\Bigg)-\lambda \sin^2\theta \mathcal{P}(\theta)=0
\end{eqnarray}
Here $\lambda$ is the separation constant and the angular equation \eqref{angular} can be solved using Associated Legendre Polynomial. For example, for a special case when $m_1=m_2=m$, \eqref{angular} will be solved by the Associated Legendre Polynomial $P_l^m(\cos{(2\theta)}$ with $\lambda =4 l(l+1)$.
\subsection{Heun equation}
Under the following changes of variables
\begin{eqnarray}
z &=& \frac{r}{r^2+r_+^2+1},\\
\Psi_{\omega} (r) &=& \left(r^3 f(r)\frac{dz}{dr}\right)^{-1/2}\mathcal{R}_{\omega}(z),
\end{eqnarray}
the radial equation \eqref{radial} can be recast as a Heun equation
\begin{eqnarray}
\label{eq:rotatingheun}
\Bigg( \partial_z^2+\frac{\frac{1}{4}-a_1^2}{(z-1)^2}-\frac{\frac{1}{2}-a_0^2-a_1^2 a_z^2+a_{\infty}^2+u}{z(z-1)}+\frac{\frac{1}{4}-a_z^2}{(z-z_0)^2}+\frac{u}{z(z-z_0)}+\frac{\frac{1}{4}-a_0^2}{z^2}\Bigg) \mathcal{R}_{\omega }(z)=0.\nonumber\\
\end{eqnarray}
Where the parameters are
\begin{eqnarray}
\label{eq:rota0}
a_0 &=& 0,\\
a_1 &=& \frac{\Delta-2}{2} ,\\
a_z &=& i\frac{\omega}{2}\frac{r_+\sqrt{1-\frac{2\alpha m_1}{\omega r_+^2}}}{1+2r_+^2},\\
a_{\infty} &=& \frac{\omega}{2}\frac{\sqrt{1+r_+^2+\frac{2\alpha m_1}{\omega}}}{1+2r_+^2},
\end{eqnarray}
\begin{eqnarray}
u = -\frac{\lambda + 2(2r_+^2+1)+r_+^2\Delta(\Delta-4)}{4(1+r_+^2)}+\frac{\alpha m_1 \omega}{2(1+r_+^2)}+\frac{r_+^2}{1+r_+^2}\frac{\omega^2}{4(2r_+^2+1)},
\end{eqnarray}
and the horizon $r_+$ is mapped to 
\begin{eqnarray}
\label{eq:rotz0}
z_0 &=& \frac{r_+^2}{2r_+^2+1}.
\end{eqnarray}
In the large black hole limit $r_+\gg 1$, if the angular velocity is small enough such that $\Omega\ll1$, the above parameters are same as those of Schwarzschild black hole when $\lambda= l(l+2)$. 
The Heun equation~\eqref{eq:rotatingheun} has four regular singularites at $z=0, 1, z_0$ and $\infty$.
\subsection{Retarded Green's function}
Following the same procedure as we did in the last section for the charged black hole, we compute the exact retarded Green's function for the rotating case as
\begin{eqnarray}
 G_R(\omega, \ell) = (1 + r_+^2)^{2a_1} e^{-\partial_{a_1}F} \frac{\sum_{\theta=\pm} \mathcal{M}_{-\theta}(a_{z}, a; a_0)\mathcal{M}_{(-\theta)+}(a, a_1; a_{\infty})  z_0^{\theta a}e^{-\frac{\theta}{2}\partial_a F} }{\sum_{\theta=\pm} \mathcal{M}_{-\theta}(a_{z}, a; a_0)\mathcal{M}_{(-\theta)-}(a, a_1; a_{\infty})  z_0^{\theta a}e^{-\frac{\theta}{2}\partial_a F}}
\end{eqnarray}
where the parameters $z_0, a_0, a_1, a_z, a_\infty$ and $u$ are written in terms of $\omega, m_1, \lambda$ and the black hole parameters $r_+$ and $\alpha$ as in~\eqref{eq:rota0}-\eqref{eq:rotz0}.
\section{Conclusion}\label{sec:con}

We have computed exact retarded Green's functions in the thermal $CFT$ with a chemical potential and that in the thermal $CFT$ with angular momenta holographically by considering Reissner-Nordstr\"{o}m and Myers-Perry black hole solutions in $AdS$ as the bulk theories respectively. We mapped the bulk field equations to the Heun equations by making appropriate changes of variables and a field redefinition. The Heun equation is mapped to the semi-classical BPZ equation satisfied by the five point correlators with degenerate insertion in the Liouville $CFT$. Using the crossing relations of the $CFT$ correlators we find the connection formulae for the solutions of the Heun equation of our interest near the regular singularities. From this connection formulae we read out the source and the response function for the bulk field and using the holographic prescription~\cite{Son:2002sd} we obtain the retarded Green's function for the scalar primary operators in the 4-dimensional thermal $CFT$.

The obvious next step would be to compute higher-point correlation functions in thermal $CFT$s. It would be interesting if this technique can be used to compute the out-of-time-ordered-correlators ($OTOC$) in the thermal $CFT$s. So far the exact retarded Green's functions are obtained for the 4-dimensional thermal $CFT$s. It would be very interesting if we can generalise the formalism to the d-dimensional case. 

We would like to compute current-current correlators, particularly $\langle J_\mu J_\nu \rangle$ and $\langle T_{\mu\nu} T_{\rho\sigma} \rangle$ in the thermal $CFT$s using this technique in future. These correlators are very useful to study transport phenomena and hydrodynamics.

We would like to interpret our results for two-point retarded Green's functions computed in the RN-$AdS$ and the rotating $AdS$ black hole backgrounds as heavy-light four point functions of zero temperature $CFT$ and solve for the heavy-light light-cone conformal bootstrap~\cite{Bonelli:2022ten,Fitzpatrick:2012yx,Komargodski:2012ek,Kulaxizi:2019tkd,Karlsson:2019dbd,Karlsson:2020ghx,Li:2019zba,Li:2020dqm}. It will also be very interesting to explore more in this direction to understand the underlying relations of 2d Liouville $CFT$ with the higher dimensional thermal $CFT$ precisely.

\section*{Acknowledgments} The work of AB is supported by the National Institute for Theoretical and Computational Sciences, NRF Grant Number 65212. The work of TM is supported by a Simons Foundation Grant Award ID 509116 and by the South African Research Chairs initiative of the Department of Science and Technology and the National Research Foundation.
\appendix 

\bibliographystyle{jhep}
\bibliography{bib_thermal}

\end{document}